\documentclass[PRB,twocolumn,groupedaddress,nofootinbib,superscriptaddress]{revtex4-1}
\usepackage{graphicx}
\usepackage{dcolumn}
\usepackage{bm}
\usepackage{times}
\newcommand{\crcl}{CrCl$_3$}

\begin{document}

\title{Magnetic Behavior and Spin-Lattice Coupling in Cleavable, van der Waals Layered CrCl$_3$ Crystals}

\author{Michael A. McGuire}
\email{McGuireMA@ornl.gov}
\affiliation{Materials Science and Technology Division, Oak Ridge National Laboratory, Oak Ridge, Tennessee 37831 USA}
\author{Genevieve Clark} \affiliation{Department of Materials Science and Engineering, University of Washington, Seattle, Washington, 98195, USA}
\author{Santosh KC}  \affiliation{Materials Science and Technology Division, Oak Ridge National Laboratory, Oak Ridge, Tennessee 37831 USA}
\author{W. Michael Chance} \affiliation{Materials Science and Technology Division, Oak Ridge National Laboratory, Oak Ridge, Tennessee 37831 USA} \affiliation{Department of Materials Science and Engineering, The University of Tennessee, Knoxville, TN 37996, USA}
\author{Gerald E. Jellison, Jr.} \affiliation{Materials Science and Technology Division, Oak Ridge National Laboratory, Oak Ridge, Tennessee 37831 USA}
\author{Valentino R. Cooper} \affiliation{Materials Science and Technology Division, Oak Ridge National Laboratory, Oak Ridge, Tennessee 37831 USA}
\author{Xiaodong Xu} \affiliation{Department of Materials Science and Engineering, University of Washington, Seattle, Washington, 98195, USA} \affiliation{Department of Physics, University of Washington, Seattle Washington, 98195, USA}
\author{Brian C. Sales} \affiliation{Materials Science and Technology Division, Oak Ridge National Laboratory, Oak Ridge, Tennessee 37831 USA}

\begin{abstract}
CrCl$_3$ is a layered insulator that undergoes a crystallographic phase transition below room temperature and orders antiferromagnetically at low temperature. Weak van der Waals bonding between the layers and ferromagnetic in-plane magnetic order make it a promising material for obtaining atomically thin magnets and creating van der Waals heterostructures. In this work we have grown crystals of CrCl$_3$, revisited the structural and thermodynamic properties of the bulk material, and explored mechanical exfoliation of the crystals. We find two distinct anomalies in the heat capacity at 14 and 17\,K confirming that the magnetic order develops in two stages on cooling, with ferromagnetic correlations forming before long range antiferromagnetic order develops between them. This scenario is supported by magnetization data. A magnetic phase diagram is constructed from the heat capacity and magnetization results. We also find an anomaly in the magnetic susceptibility at the crystallographic phase transition, indicating some coupling between the magnetism and the lattice. First principles calculations accounting for van der Waals interactions also indicate spin-lattice coupling, and find multiple nearly degenerate crystallographic and magnetic structures consistent with the experimental observations. Finally, we demonstrate that monolayer and few-layer CrCl$_3$ specimens can be produced from the bulk crystals by exfoliation, providing a path for the study of heterostructures and magnetism in ultrathin crystals down to the monolayer limit.

\end{abstract}

\maketitle

\section{Introduction}

Recent interest in layered ferromagnetic materials is driven by the desire to develop functional van der Waals heterostructures \cite{Geim-2013, Zhong-2017}. For this purpose materials must be cleavable down to very thin, ideally monolayer, specimens. This is possible when the layers composing the crystal are held together by weak van der Waals bonds so they can be mechanically separated or exfoliated by chemical intercalation routes \cite{Nicolosi-2013}. This led to the identification and study of CrSiTe$_3$ and CrGeTe$_3$ \cite{Lebegue-2013, Li-2014, Casto-2015, Sivadas-2015}. These are small band gap semiconductors with Curie temperatures of 33\,K for CrSiTe$_3$ \cite{Carteaux-1991} and 61 K for CrGeTe$_3$ \cite{Carteaux-1995}. The cleavage energy of these materials is calculated to be 0.35-0.38 J/m$^2$ \cite{Li-2014}, similar to graphite (0.43 J/m$^2$) and MoS$_2$ (0.27 J/m$^2$) \cite{Bjorkman-2012}. Studies of nanosheets of CrSiTe$_3$ suggest ferromagnetism may persist in ultrathin specimens \cite{Lin-2016}, and ferromagnetic few-layer-thick crystals of CrGeTe$_3$ have recently been reported \cite{Gong-2017}. The tin analogue CrSnTe$_3$ is also predicted to be a ferromagnet \cite{Zhuang-2015}. Other recently discovered materials include Fe$_3$GeTe$_2$ \cite{Deiseroth-2006}, a metal with itinerant ferromagnetism below $T_C$=220-230 K \cite{Deiseroth-2006, Chen-2013, May-2016}. Chromium triiodide, and transition metal halides in general \cite{McGuire-Review}, have also been put forth as candidates for cleavable magnetic materials \cite{McGuire-2015, Zhang-2015, Liu-2016, Wang-2016}. Experimentally, CrI$_3$ is a ferromagnet with $T_C$\,=\,61\, K \cite{Hansen-1959, Dillon-1965, McGuire-2015}, CrBr$_3$ is a ferromagnet with $T_C$ = 37 K \cite{Tsubokawa-1960}, and CrCl$_3$ is an antiferromagnet with an ordering temperature near 17 K \cite{Hansen-1958, Cable-1961}. Several theoretical studies have been recently published on chromium trihalides addressing bulk magnetic properties and behavior in monolayer form \cite{Wang-2011, McGuire-2015, Zhang-2015, Liu-2016, Wang-2016, Lado-2017}, and also on VCl$_3$ and VI$_3$, which are predicted to be ferromagnetic and Dirac half-metals \cite{He-2016}. All of the calculations predict low cleavage energies for these compounds, similar to those noted above for graphite and MoS$_2$. Recently ferromagnetism has been experimentally confirmed in monolayer CrI$_3$ with intriguing thickness dependent magnetic phases \cite{Huang-2017}, and heterostructures incorporating ultrathin CrI$_3$ have enabled remarkable control of the spin and valley pseudospin properties in monolayer WSe$_2$ through a large exchange field effect \cite{Zhong-2017}.

It is interesting to note that the magnetic ordering temperatures of the chromium trihalides increase as the halogen size increases from Cl to Br to I. Since the Cr-Cr distances increase with increasing halogen size, the direct exchange is expected to be weakened along this series. This indicates that superexchange, which is expected to favor ferromagnetic alignment, is the more important magnetic interaction \cite{Lado-2017}. As the electronegativity is decreased from Cl to Br to I, it is expected that the Cr-halogen bonding becomes more covalent, strengthening superexchange interactions and raising ordering temperatures. Indeed all three of these chromium trihalides exhibit ferromagnetic ordering of Cr moments within the layers at low temperature \cite{deHaas-1940, Cable-1961, Tsubokawa-1960, McGuire-2015}. Moving from Cl to Br to I is also expected to increase spin orbit coupling associated with the halogen, which has been identified as a source of magnetic anisotropy in these materials \cite{Lado-2017}.

The present work focuses on chromium trichloride. Initial studies of magnetism in this material date back nearly 100 years \cite{Woltjer-1925}. The antiferromagnetic ground state consists of moments lying in the plane defined by the Cr layers. Those moments are aligned ferromagnetically within a layer, and the layers stack antiferromagnetically, as demonstrated by neutron diffraction \cite{Cable-1961}. The transition temperature of 17\,K  was identified by heat capacity measurements \cite{Hansen-1958, Starr-1940}. Kuhlow performed Faraday rotation measurements and concluded that upon cooling the magnetic order appeared to develop by first forming two-dimensional ferromagnetic order within the layers and then, at slightly lower temperature, forming long range three-dimensional order through antiferromagnetic coupling between the layers \cite{Kuhlow-1982}. Faraday rotation, magnetization, and neutron diffraction measurements show that the ordered state has very little anisotropy, and fields of only a few kOe are required to fully polarize the magnetization in or out of the plane \cite{Bizette-1961, Cable-1961, Kuhlow-1982}. Weak magnetic anisotropy and weakly antiferromagnetic interlayer interactions are also reported from spin wave analysis \cite{Narath-1965}. In addition, CrCl$_3$ is known to undergo a crystallographic phase transition near 240\,K, similar to CrBr$_3$ and CrI$_3$, corresponding to a change in the layer stacking arrangement and a transition from monoclinic ($C2/m$) at high temperature to rhombohedral ($R\overline{3}$) at low temperature with little change in the intralayer structure \cite{Morosin-1964, McGuire-2015}.

There have been several recent theoretical studies of CrCl$_3$ based on first principles electronic structure calculations. Wang \textit{et al.} employed all-electron calculations for bulk CrCl$_3$, and the experimentally observed rhombohedral-antiferromagnetic ground state was reproduced for one of the methods used for incorporating U, the on-site Coulomb repulsion for Cr \cite{Wang-2011}. The authors noted that the energetics of the different crystal and magnetic structures are sensitive enough to the specific approach used that it is difficult to draw definitive conclusions about their relative stability, consistent with the experimentally observed temperature induced crystallographic phase transition and the weak magnetic anisotropy in the antiferromagnetic state. Other theoretical studies focus mainly on CrCl$_3$ monolayers. Liu \textit{et al.} observed ferromagnetic order below 66\,K in Monte Carlo simulations of monolayers, and suggested that hole doping should increase the Curie temperature \cite{Liu-2016}. The cleavage energy in the bulk crystal was calculated using a van der Waals density functional (vdW-DF) method to account for the dispersion forces in the bulk crystal. This gave a very low cleavage energy of 0.10$-$0.13\,J/m$^2$. Zhang \textit{et al.} calculate a cleavage energy of 0.3\,J/m$^2$ using a different vdW-DF \cite{Zhang-2015}. Monte Carlo simulations using their first principles calculations results for monolayer CrCl$_3$ suggest a Curie temperature of 49\,K, which is predicted to increase with applied strain. The results of those two studies indicate that CrCl$_3$ monolayers should be mechanically stable.

Here we report results from a thorough study of bulk \crcl\ crystals, including x-ray diffraction, magnetization, and heat capacity data, along with van der Waals density functional studies of the structure and magnetism. In addition we include optical images and atomic force microscopy measurements on ultrathin specimens cleaved from the bulk crystals. Our observations for the bulk crystals are in agreement with previously published literature, and in addition: (1) we note an anomaly in the magnetic susceptibility at the crystallographic phase transition indicating some coupling of the magnetism to the crystal lattice, (2) we find good structural agreement between theory and experiment only when magnetism is included in the calculations, again indicating spin-lattice coupling, (3) we obtain an estimate of the in-plane spin-flop field from magnetization measurements, (4) we observe two separate heat capacity anomalies associated with magnetic ordering indicative of short-ranged or two-dimensional ferromagnetism evolving into long-range antiferromagnetism, (5) we present the evolution of isothermal magnetization curves that is consistent with this scenario and construct a temperature-field phase diagram for CrCl$_3$, and (6) we demonstrate stable monolayer and few-layer specimens can be cleaved from bulk crystals. Overall, it is apparent that CrCl$_3$ is a promising material for the study of magnetism in ultrathin crystals and for incorporating magnetism into van der Waals heterostructures.


\section{Procedures}

Crystal growth is described in the following Section. A PANalytical X-Pert Pro MDP diffractometer equipped with an Oxford PheniX cryostat was used for x-ray diffraction measurements (Cu-K$_{\alpha1}$ radiation). Heat capacity and ac magnetization measurements were performed using a Quantum Design PPMS, and dc magnetization was measured using a Quantum Design MPMS. Optical absorbance \textit{A} was determined from the measured optical transmittance \textit{T} by $A = -log(T)$. Transmission measurements were made using a home-built system specifically designed for small samples.  The light source was an Energetiq LDLS 99, which was collimated coming from the fiber. The detector was an Ocean Optics USB4000 0.1 meter spectrometer attached to a 600 micron multimode fiber.  The sample was placed directly in front of the 600 micron entrance of the fiber.  The transmission was determined by the ratio of the spectroscopic intensity with the sample in place and the intensity with the sample removed. Exfoliation of bulk CrCl$_3$ was achieved using mechanical exfoliation with scotch tape onto 90\,nm SiO$_2$ substrates. Optical microscopy images were obtained using an Olympus BX51M microscope, while atomic force microscopy (AFM) was carried out using a Bruker Edge Dimension atomic force microscope.

DFT calculations were performed using the Vienna Ab-initio Simulation Package (VASP) \cite{Kresse-1996} with the projector-augmented wave (PAW) \cite{Blochl-1994} potentials in order to understand the crystallographic and electronic properties of bulk CrCl$_3$. The exchange-correlation was approximated with generalized gradient approximation (GGA) of Perdew-Burke-Ernzerhof (PBE) functionals \cite{Perdew-1996} as well as van der Waals correction using vdW-DF-optB86b functional \cite{Dion-2004, Klimes-2010, Thonhauser-2007, Thonhauser-2015}. The pseudopotentials used explicitly treat $3p^6$, $4s^1$, $3d^5$, and $3s^2$, $3p^5$ electrons as valence electrons for Cr and for Cl, respectively. The Brillouin zone (BZ) integration was performed using the Monkhorst-Pack \cite{Monkhorst-1976} sampling method with a 4$\times$4$\times$2 ($R\overline{3}$)  and 4$\times$2$\times$2 ($C2/m$)  k-meshes for structural optimization.The energy cutoff was 500\,eV and the criteria for energy and force convergence were set to be $1\times10^{-4}$\,eV and 0.01\,eV/{\AA}, respectively.


\section{Results and Discussion}

\begin{figure}
\begin{center}
\includegraphics[width=3.0in]{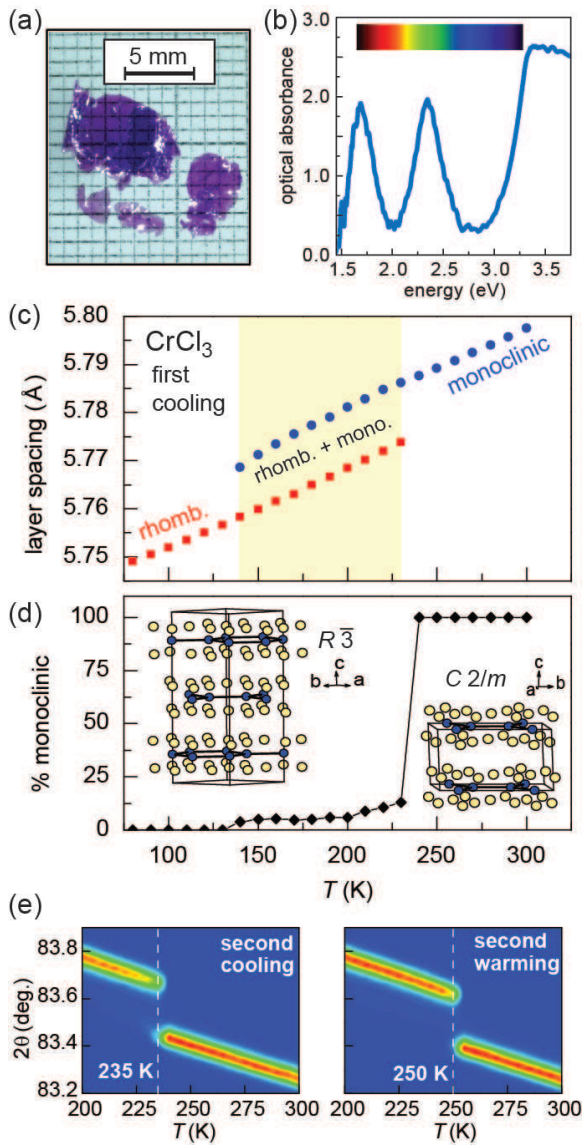}
\caption{\label{fig:crystals}
(a) CrCl$_3$ single crystals grown for this study. (b) The optical absorbance spectrum showing strong absorption of green and red resulting in the observed violet color. (c) Temperature dependence of the layer spacing in CrCl$_3$ measured using x-ray diffraction during the first cooling of the crystal. (d) Relative amount of the monoclinic phase present during the first cooling through the monoclinic to rhombohedral phase transition. The two crystal structures are depicted on this panel as well. (e) Contour plot of the diffracted intensity from the monoclinic 0\,0\,5 and rhombohedral 0\,0\,15 reflections on the second cooling and warming cycles.
}
\end{center}
\end{figure}

\subsection{Crystal growth, structure, and spin-lattice coupling}

Thin, plate-like single crystals of CrCl$_3$ with lateral dimensions up to several millimeters are easily grown by recrystallizing commercial CrCl$_3$ using chemical vapor transport \cite{Starr-1940}. Here CrCl$_3$ from Alpha Aesar with a metals-basis purity of 99.9\,\% was used. The as-received material, in the form of small platelets, was sealed inside evacuated silica tubes. A typical growth used about 1\,g of CrCl$_3$ in a 15\,cm long tube with 16\,mm inner diameter and 1.5\,mm wall thickness. The tubes were placed in a horizontal tube furnace so that the starting material was at the end of the tube near the center of the furnace and the other end of tube was near the opening at the end of the furnace. The temperature at the center of the furnace was set to 700\,$^{\circ}$C, and the temperature at the cooler end of the tube was measured to be 550\,$^{\circ}$C. The growth begins rapidly and slows as the starting material is consumed. Large, violet-colored, transparent platelets like those shown in Fig. \ref{fig:crystals}a are present after 48\,h at temperature, while it may take up to a week for all of the CrCl$_3$ to be transported. After the recrystallizations are complete a green powder, presumably Cr$_2$O$_3$, is left at the hot end of the tubes. This is attributed to oxide impurity in the starting material.

The optical absorbance of a CrCl$_3$ crystal grown for the present study is shown in Fig. \ref{fig:crystals}b. A color scale showing the approximate color of the visible light for photon energies between 1.7 and 3.3\,eV is included on the plot. The results are consistent with the data reported in the thorough study of the optical properties of this material by Pollini and Spinolo \cite{Pollini-1970}. A band gap of 3.1\,eV is approximated by the onset of strong absorption at higher energies. Below the band edge, there are two broad absorptions in the red and green centered near 1.7 and 2.3\,eV, to which the violet color of the crystal can be attributed. These bands, which have fine structure not resolved here, arise from Cr$^{3+}$ \textit{d}-\textit{d} transitions as described in detail in Ref. \cite{Pollini-1970}.

As noted above, CrCl$_3$ is known to undergo a crystallographic phase transition below room temperature. Using single crystal diffraction, Morosin and Narath \cite{Morosin-1964} demonstrated that CrCl$_3$ adopts the monoclinic space group $C2/m$ at room temperature and the rhombohedral space group $R\overline{3}$ at 225\,K. They used nuclear quadrupole resonance to identify the phase transformation temperature as 240\,K. The crystallographic phase transitions involves mainly a change in the layer-to-layer shift in the stacking sequence. The intralayer structure remains nearly the same, with an ideal honeycomb net of Cr in the rhombohedral structure and a slightly distorted honeycomb net in the monoclinic structure \cite{Morosin-1964, McGuire-2015}. For example, in CrCl$_3$ the in-plane Cr-Cr distances are 3.431\,{\AA} at 225\,K (rhombohedral) and 3.440 and 3.441\,{\AA} at 298\,K (monoclinic) \cite{Morosin-1964}.

Results of x-ray diffraction measurements through this phase transition are shown in Figure \ref{fig:crystals}c-e. During the first cooling the high and low temperature phases coexist over a relatively wide temperature range (Fig. \ref{fig:crystals}c). The fraction of the monoclinic phase present during this initial cool down, determined by the ratio of the peak intensities, is shown in Figure \ref{fig:crystals}d. More than 85\% of the crystal transforms between 240 and 230\,K, and the remaining small fraction of the high temperature phase essentially vanishes between 140 and 130\,K. The two crystal structures are shown on this figure.

During subsequent cooling and heating cycles the phase transformation is much sharper (Fig. \ref{fig:crystals}e), as was found to the case in CrI$_3$ crystals \cite{McGuire-2015}. This suggests that the retained high temperature phase present in the first cool down is likely associated with defects or strain frozen into the crystals during the growth at elevated temperature and subsequent cooling to room temperature. There is still significant thermal hysteresis after the first thermal cycling, consistent with the first order nature of the phase transition.

\begin{table*}
\begin{center}
\caption{\label{tab:dft}
Structural parameters from DFT calculations using PBE/GGA and vdW-DF-optB86b for bulk CrCl$_3$ in the monoclinic and rhombohedral crystal structures, with corresponding experimental values from Ref. \citenum{Morosin-1964}. Energies given in eV/atom can be converted to eV/Cr by multiplying by four.}
\setlength{\tabcolsep}{2mm}
\begin{tabular}{c|cc|cc|cc|c}
\toprule
\multicolumn{8}{c}{Monoclinic,  $C2/m$}															\\
\hline															
	&	 \multicolumn{2}{c|}{non-magnetic}			&	 \multicolumn{2}{c|}{ferromagnetic}			&	 \multicolumn{2}{c|}{antiferromagnetic}			&	exp. \cite{Morosin-1964}	\\
	&	PBE	&	vdW-DF	&	PBE	&	vdW-DF	&	PBE	&	vdW-DF	&	$T = 298$\,K	\\
\hline															
a ({\AA})	&	5.992	&	5.825	&	6.060	&	5.944	&	6.039	&	5.945	&	5.959	\\
b ({\AA})	&	10.372	&	9.404	&	10.490	&	10.288	&	10.457	&	10.290	&	10.321	\\
c ({\AA})	&	6.517	&	6.086	&	6.656	&	6.043	&	6.571	&	6.036	&	6.114	\\
$\beta$ (deg.)	&	107.3	&	107.5	&	107.1	&	108.6	&	107.3	&	108.7	&	108.5	\\
vdW gap ({\AA})	&	3.775	&	3.100	&	3.869	&	3.245	&	3.772	&	3.205	&	3.299	\\
Energy (eV/atom)	&	--	&	-3.503	&	--	&	-3.785	&	--	&	-3.785	&	--	\\
\toprule															
\multicolumn{8}{c}{Rhombohedral,  $R\overline{3}$}															\\
\hline															
	&	 \multicolumn{2}{c|}{non-magnetic}			&	 \multicolumn{2}{c|}{ferromagnetic}			&	 \multicolumn{2}{c|}{antiferromagnetic}			&	exp. \cite{Morosin-1964}	\\
	&	PBE	&	vdW-DF	&	PBE	&	vdW-DF	&	PBE	&	vdW-DF	&	$T = 225$\,K	\\
\hline															
a ({\AA})	&	5.652	&	5.563	&	6.054	&	5.942	&	6.053	&	5.930	&	5.942	\\
c ({\AA})	&	19.641	&	17.288	&	19.239	&	17.088	&	19.027	&	17.184	&	17.333	\\
vdW gap ({\AA})	&	3.816	&	3.035	&	3.746	&	3.037	&	3.676	&	3.070	&	3.093	\\
Energy (eV/atom)	&	--	&	-3.454	&	--	&	-3.786	&	--	&	-3.786	&	--	\\
\toprule
\end{tabular}
\end{center}
\end{table*}

Results of our DFT calculations for CrCl$_3$ in both the monoclinic and rhombohedral structures are summarized in Table \ref{tab:dft}. We investigated non-magnetic (NM), ferromagnetic (FM), and antiferromagnetic (AFM) configurations for both crystal structures. The AFM configuration refers to the experimental magnetic structure, with ferromagnetic planes stacked antiferromagnetically. Results are presented for both PBE and vdW-DF-optB86b functionals. For one of these cases, the FM rhombohedral structure, results have been previously reported by Zhang \textit{et al.} using PBE and the van der Waals density functional optB88 \cite{Zhang-2015}.The results shown in Table \ref{tab:dft} for this configuration are within 1\% of the values reported there. Our calculations indicate the monoclinic structure to be favored by 0.05\,eV/atom in the non-magnetic calculations, and essentially degenerate monoclinic and rhombohedral structures for the magnetically ordered cases, with a very small 0.001\,eV/atom preference for the experimentally observed rhombohedral structure. The presence of nearly degenerate crystallographic ground states is consistent with the observed temperature induced structural phase transition.

The cleavage energy calculated in the rhombohedral structure with vdW-DF-optB86B is 0.3\,J/m$^2$, in agreement with the calculated value for CrCl$_3$ given in Ref. \citenum{Zhang-2015}. This is similar to values determined using similar approaches for the other chromium trihalides \cite{Zhang-2015, McGuire-2015}, as well as other related, easily-cleaved materials noted in the Introduction. The low cleavage energy enables monolayer and few-layer specimens to be cleaved from the CrCl$_3$ crystals, as demonstrated below.

The optimized lattice parameters for the FM and AFM unit cells are similar to one another, and those determined from vdW-DF-optb86b calculations agree well with the experimental structure (Table \ref{tab:dft}). As expected the PBE results show an elongation along the stacking directions, the \textit{c} axes. It is particularly interesting to note the significant difference between the optimized lattice parameters for the NM structure and those determined for the magnetically ordered structures. In each case the NM unit cell is significantly smaller. This is true for both the monoclinic and rhombohedral structures. For the vdW-DF-optb86b results in the AFM and FM structures, the optimized unit cell volumes are within 2\% of the experimental values. For the NM case they are smaller by more than 10\% (Table \ref{tab:dft}). The reason for this is not completely clear, but it does suggest a strong coupling between the spins and the crystal lattice in CrCl$_3$. The difference in unit cell volume arises primarily from variation in the in-plane lattice constants, where magnetic exchange is expected to be the strongest; however, it is important to note that the method used here includes spin only in the local part of the exchange-correlation functional, and not in the non-local part \cite{Thonhauser-2015}, which may be more important in the interplanar interactions. Crystallographic studies through the magnetic ordering temperature would be helpful in probing the spin-lattice coupling identified in these calculations. Although our diffraction measurements do not extend to low enough temperature to examine this in CrCl$_3$, a structural response at the magnetic ordering temperature was noted in isostructural CrI$_3$ \cite{McGuire-2015}.

\begin{figure*}
\begin{center}
\includegraphics[width=6.5in]{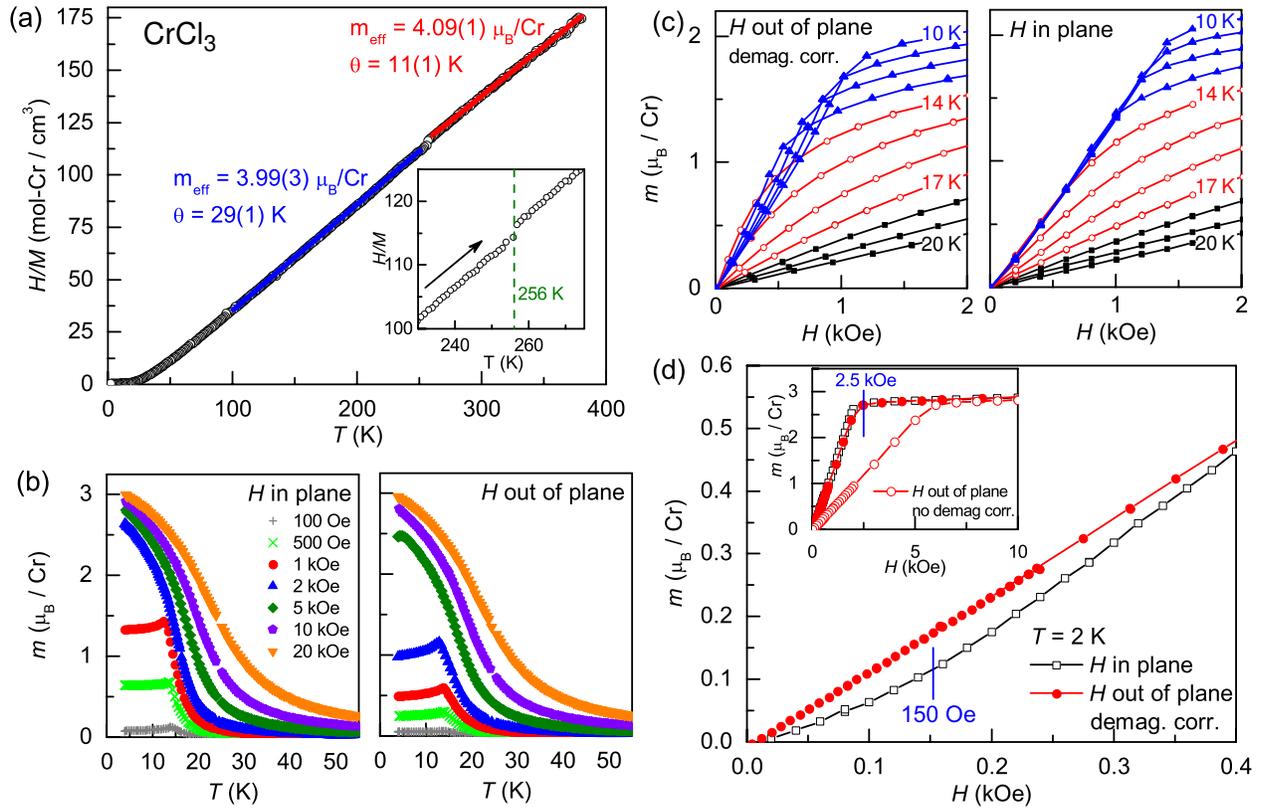}
\caption{\label{fig:MT}
Measured magnetic behavior of CrCl$_3$ crystals. (a) Temperature dependence of the inverse susceptibility (H/M) measured with a 10\,kOe field applied in the plane showing the small anomaly at the structural phase transition (inset) and Curie-Weiss fits to data above and below the transition. (b) Moment per Cr atom vs temperature near the magnetic ordering measured upon cooling in the indicated applied magnetic fields both in and out of the plane. (c) Isothermal magnetization curves near the magnetic ordering temperatures. Curves were measured at temperature intervals of 1\,K and data sets at 10, 14, 17, and 20\,K are labeled. (d) Moment per Cr atom measured at 2\,K. Due to the thin plate-like shape of the crystals, a demagnetization factor of 4$\pi$ (cgs units) was used to determine the internal field for the field out of the plane measurements. The inset shows both uncorrected (open circles) and corrected data (solid circles).
}
\end{center}
\end{figure*}

Additional evidence of spin-lattice coupling is seen in the magnetization data near the structural phase transition plotted in Fig. \ref{fig:MT}a. Temperature dependent magnetic susceptibility data, plotted as inverse susceptibility $H/M$, reveal a small discontinuity at 256\,K. The data was collected on warming, and the temperature of the discontinuity agrees well with the transition temperature determined by x-ray diffraction on warming (Fig. \ref{fig:crystals}e). Similar evidence of coupling between the magnetism and crystal lattice was previously noted for CrI$_3$ \cite{McGuire-2015}. Above and below this temperature Curie-Weiss behavior is observed, down to about 100\,K and up to 380\,K, the highest temperature investigated here. The results are in agreement with other analyses of high temperature data \cite{Starr-1940, Hansen-1959a}, although the anomaly at the structural phase transition was not noted in previous measurements due to relatively sparse data. Both the high and low temperature fits in Fig. \ref{fig:MT}a give an effective moment that is close to the expected value for spin-only Cr$^{3+}$ (3.87\,$\mu_B$).


The primary difference between the paramagnetic behavior above and below the structural transition is in the Weiss temperature $\theta$, which is smaller in the high temperature monoclinic phase than in the low temperature rhombohedral phase. The Weiss temperatures are positive, indicating that the intraplanar ferromagnetic interactions dominate the magnetic behavior in the paramagnetic state. The strength of the magnetic interactions, primarily superexchange, are expected to be sensitive to the details of the chemical coordination, allowing coupling of the magnetic behavior to the crystal structure details. In particular, the antiferromagnetic interplanar superexchange interactions will depend upon how the CrCl$_3$ layers stack. Since the crystallographic phase transition amounts essentially to a change in the layer stacking, this is identified as a likely source of the coupling between the magnetism and the lattice that produces the magnetic anomaly near 250\,K.

While both CrCl$_3$ and CrI$_3$ {\cite{McGuire-2015} show evidence of spin-lattice coupling, key differences can be noted in their behaviors and magnetic structures. The most obvious difference is the orientation of the ordered moments in the two phases, in-plane for the chloride and out of plane for the iodide (and bromide). In addition, the response of the magnetic susceptibility to the structural phase transition, as quantified by fits using a Curie-Weiss model, differ in the two materials. In CrCl$_3$, the Weiss temperature changes by nearly a factor of three, while in CrI$_3$ this parameter changes little and a response is seen mainly in the effective moment. As noted above, the crystallographic layer spacing was seen to decrease upon cooling into the magnetically ordered state CrI$_3$ \cite{McGuire-2015}. A similar type of coupling between the crystal structure and magnetic order may be expected in CrCl$_3$, though the present diffraction measurements do not extend to low enough temperature to test this hypothesis.  A detailed crystallographic study of chromium trihalides through the magnetic ordering temperatures as well as a comparative theoretical study incorporating non-collinear magnetic structures and spin-orbit coupling would be desirable to help quantify and better understand the coupling between the magnetism and the lattice in these compounds.


\subsection{Magnetic behavior and phase transitions}

Magnetization data from our crystals near the magnetic ordering transitions are summarized in Figure \ref{fig:MT}b-c. The temperature dependent data were collected upon cooling, and the isothermal magnetization curves were collected upon decreasing the applied field. The results are in general agreement with previous studies employing a variety of techniques, which find below about 14\,K moments in each layer ferromagnetically aligned and lying in the \textit{ab}-plane with antiferromagnetic stacking between planes along the \textit{c}-axis and weak magnetic anisotropy \cite{Bizette-1961, Cable-1961, Narath-1965, Kuhlow-1982}. The temperature dependence of the magnetic moment per Cr atom measured near the Curie temperature is shown in Figure \ref{fig:MT}b. At 4\,K and 2\,T a moment of 3.0\,$\mu_B$ per Cr is measured, as expected for S\,=\,$\frac{3}{2}$ trivalent Cr. Data are shown for multiple applied magnetic fields (no correction for demagnetization) with the field both in the plane of the CrCl$_3$ layers and out of the plane. The curves are very similar to those reported in Ref. \citenum{Bizette-1961}. Kuhlow reported similar behavior in Faraday-rotation measurements \cite{Kuhlow-1982}. In that study, the magneto-optical measurements indicated that the magnetic order developed in two steps upon cooling, corresponding to the upturn in the measured magnetization (Fig. \ref{fig:MT}b) below about 20\,K and the sharp cusp apparent in low field measurements near 14\,K. This was interpreted as 2D ferromagnetic order within the layers developing first, with interlayer long-range antiferromagnetic order setting in at lower temperature.

This two-step development of magnetic order can also be inferred from the isothermal magnetization curves of Bizette \textit{et al}. \cite{Bizette-1961}, as pointed out by Kuhlow, who saw similar behavior in isothermal Faraday rotation curves \cite{Kuhlow-1982}. This behavior is demonstrated for the crystals used in the present study in Figure \ref{fig:MT}c. Magnetization is shown as a function of field for temperatures between 10 and 20\,K, with the field applied out of the plane defined by the CrCl$_3$ layers and with the field applied out of this plane. Data collected with the field normal to the thin platelet crystals was corrected for demagnetization effects by assuming a demagnetization factor of 4$\pi$ (cgs units). The curves in Figure \ref{fig:MT}c can be divided into three families based on their curvature, and are delineated by different colors and symbols in the figure. The distinctions are most apparent in the out of plane data. Between 20 and 18\,K the curves are linear. At 17\,K a negative curvature develops, indicating a ferromagnetic-like response. This behavior is enhanced upon cooling to 14\,K. At 13\,K the response changes to nearly linear at low fields, characteristic of the behavior expected for polarizing an antiferromagnet with low anisotropy. Thus, the data suggests the onset of ferromagnetic correlations at 17\,K and antiferromagnetic order at 13\,K, in general agreement with Kuhlow's observations \cite{Kuhlow-1982}.

Results of heat capacity measurements on our CrCl$_3$ crystals are summarized in Figure \ref{fig:hc}. The data clearly show two thermal anomalies, a broad feature centered at 17.2\,K and a sharp lambda-like peak centered at 14.1\,K. Data from two samples are shown in Figure \ref{fig:hc}a and they are nearly identical. These well-resolved anomalies are not reported in previous heat capacity studies, which only observe a broad peak near 17\,K \cite{Hansen-1958, Kostryukova-1972}. The presence of two well-separated thermal signatures in the heat capacity data at low magnetic fields shown in Figure \ref{fig:hc}a strongly supports the interpretation of the magnetic data in terms of the evolution of magnetic order proposed by Kuhlow \cite{Kuhlow-1982} and described above.

The magnetic heat capacity $c_{mag}$ was estimated by subtracting a smooth curve fitted to the zero field data at temperatures between 2 and 50\,K excluding the 7.5$-$33\,K range. Figure \ref{fig:hc}c shows results for data collected in applied magnetic fields of 0 and 5\,kOe. Integrating $c_{mag}/T$ to extract the magnetic entropy released up to 30\,K gives about 2\,J\,K$^{-1}$\,mol-Cr$^{-1}$ at both fields. This is a small fraction of the expected value of $R\rm{ln}(2S+1)$\,=\,11.5\,J\,K$^{-1}$\,mol-Cr$^{-1}$, and suggests that most of the magnetic entropy is released as magnetic correlations develop well above the onset of long range order. This is consistent with previous observations for isostructural CrBr$_3$ and CrI$_3$ \cite{Jennings-1965, McGuire-2015}.

Similarly small lambda anomalies have been reported at the long range ordering temperatures in honeycomb layered compounds, for example MnPS$_3$ \cite{Takano-2004, Wildes-2006} and BaCo$_2$(AsO$_4$)$_2$ \cite{regnault-1990}, in which magnetic fluctuations are expected to be responsible for the ``missing'' entropy. Development of these fluctuations produces a very broad contribution to the temperature dependence of the heat capacity extending to well above the ordering temperature. Thus, while significant magnetic fluctuations are expected to be present in CrCl$_3$, they are likely not the origin of the ferromagnetic-like response seen just above 14\,K, since a relatively sharp thermal anomaly is seen at the onset of that behavior.

\begin{figure}
\begin{center}
\includegraphics[width=3.0in]{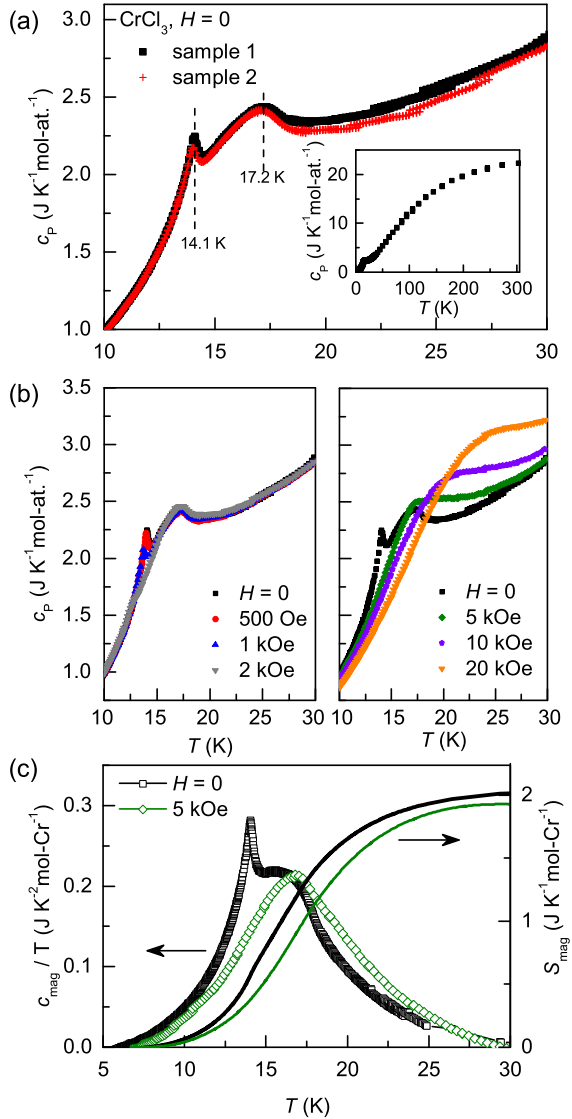}
\caption{\label{fig:hc}
Heat capacity of CrCl$_3$ crystals. (a) Data collected on two different samples near the magnetic ordering, with position of the two observed peaks noted. Data from 2 to 300 K are shown in the inset. (b) Magnetic field dependence of the heat capacity at low fields (left) and higher fields (right). The field was applied out of the plane. (c) Estimated magnetic heat capacity plotted as $c_{mag}/T$ (symbols) determined for data collected in zero field and at 20 kOe, and the associated magnetic entropies (lines).
}
\end{center}
\end{figure}

To further probe the nature of the magnetic phase transitions, heat capacity data were collected in applied magnetic fields (normal to the CrCl$_3$ layers) and the data are presented in Figure \ref{fig:hc}b. For small applied magnetic fields, up to 2\,kOe, the feature near 17\,K is nearly unchanged while the feature at 14\,K is suppressed in magnitude and temperature. At applied fields of 5\,kOe and higher the low temperature feature is completely suppressed and the 17\,K feature begins to broaden and move to higher temperature. The latter behavior is typical of heat capacity anomalies associated with ferromagnetic order, which is reinforced by an applied magnetic field.  The heat capacity anomaly at 14\,K is suppressed with field much more rapidly than is seen in typical antiferromagnets. In this case, such behavior can be attributed to the competing ferromagnetic order at low applied fields in this weakly anisotropic magnetic system.

Two heat capacity anomalies have been observed in data from the closely related compound RuCl$_3$, and have been associated with different types of magnetic order occurring in different parts of the sample that have different crystal structures or stacking fault densities \cite{Banerjee-2016, RuCl3-Cao-2016}. Several observations suggest that this is not the case for the CrCl$_3$ data shown in Fig. \ref{fig:hc}. First, identical behavior is observed in two samples selected from separate growths (Fig. \ref{fig:hc}a). Second, the evolution of the two heat capacity anomalies with applied field are coupled as described above. Third, the magnetization data (Fig. \ref{fig:MT}c) show that the FM-like behavior onsetting at 17\,K vanishes at the 14\,K transition indicating that the magnetic behavior of the entire sample is affected. Fourth, the ferromagnetic in-plane magnetic structure of CrCl$_3$ should result in less sensitivity to stacking faults than the more complex zig-zag antiferromagnetic order in RuCl$_3$.

Using heat capacity data like that shown in Figure \ref{fig:hc}b, where the local maxima can be easily identified and tracked as the applied magnetic field is changed, the magnetic phase diagram for CrCl$_3$ shown in Figure \ref{fig:PD} can be constructed. Three regions are delineated: (1) the paramagnetic (PM) phase at high temperatures and low fields, (2) the antiferromagnetic (AFM) region where the magnetic moments are aligned ferromagnetically within layers and the layers are stacked antiferromagnetically, and (3) a ferromagnetic-like phase (``FM''), which is either field polarized in the case of the high field region or short ranged or low dimensional in the case of the low field region between the PM and AFM phases. Note that the boundaries are not as well defined as this depiction might suggest, especially at higher magnetic fields.

\begin{figure}
\begin{center}
\includegraphics[width=3.25in]{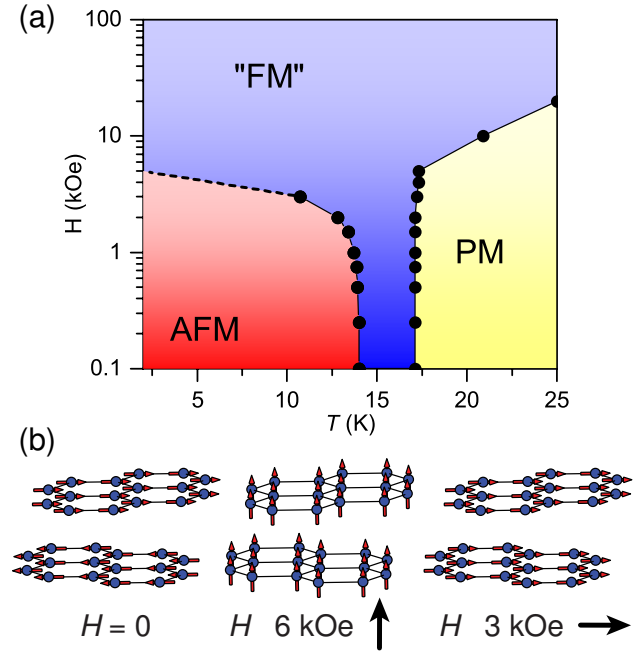}
\caption{\label{fig:PD}
(a) Magnetic phase diagram of CrCl$_3$ crystals. $H$ is the applied external field, directed out of the plane of the Cr layers. The points were determined from heat capacity measurements like those shown in Figure \ref{fig:hc}b. PM = paramagnetic, AFM = antiferromagnetic, ``FM'' = ferromagnetic-like. (b) Magnetic structures of CrCl$_3$ accessible with relatively small applied magnetic fields.
}
\end{center}
\end{figure}
%


\subsection{Magnetic anisotropy and potential for monolayer and heterostructure studies}

As noted above, the observed magnetic anisotropy in the ordered state is small. Figure \ref{fig:MT}d shows the magnetic moment as a function of magnetic field measured at 2\,K with the field in the plane and out of the plane. The out of plane data is shown both as a function of internal field (demag. corr.) and applied field (no demag. corr.) and is consistent with the data of Narath and Davis \cite{Narath-1965}.  Demagnetization effects are significant as expected. An applied field of 6\,kOe is required to essentially fully polarize the magnetization out of the plane, consistent with neutron diffraction results \cite{Cable-1961}. This corresponds to an internal field of only 2.5\,kOe. Interestingly, this is the same field required to polarize the magnetization in the plane. Since within each plane the moments are ferromagnetically aligned and lie in the plane, it is expected that polarization out of the plane is realized through a coherent rotation of moments within each layer \cite{Kuhlow-1982}. When the field is applied in the plane, different field induced behaviors are expected depending on whether the field is parallel or perpendicular to the moment direction. In the latter case, a coherent rotation occurs, while in the former case a spin flop can be expected to occur first. The sample used for the in plane measurement comprised several thin crystals that were not co-aligned with respect to the \textit{ab}-plane, so the direction of the applied field relative to the moment direction in the zero field magnetic structure is not well defined. As a result the signature of the spin flop on the magnetization data is somewhat diluted, but it is still observable in Figure \ref{fig:MT}d. The spin flop is seen to occur at a field between 100 and 200\,Oe at \textit{T}\,=\,2\,K, consistent with the value of 163\,Oe estimated from magneto-optical measurements at 7\,K from which an in-plane anisotropy field of about 10\,Oe is determined \cite{Kuhlow-1982}. This is small relative to the field required to rotate the moments out of the plane, suggesting an XY-Heisenberg description for the moments in CrCl$_3$.

Note that the energies of the FM and AFM configurations in Table \ref{tab:dft} are the same within the precision of the calculations; the magnetically ordered structures are about 0.3\,eV/atom lower in energy than the NM states in both the monoclinic and rhombohedral crystal structures in the vdW-DF-optb86b calculations. This is similar to the 0.4\,eV/atom stability found in the calculations of Ref. \citenum{Wang-2011}. This appears consistent with the observation that the moments can be fully polarized with relatively small applied fields. Note that spin-orbit coupling, a key source of magnetocrystalline anisotropy, is not included in the calculations. However, this effect is expected to be weak in trivalent chromium with electronic configuration $3d^3$ and nearly octahedral coordination due to quenching of the orbital moment. Thus single ion anisotropy should be weak for Cr, as was demonstrated recently for CrI$_3$ by Lado and Fern{\'{a}}ndez-Rossier \cite{Lado-2017}. In that study, spin-orbit coupling on the heavy iodine ions was found to contribute the majority of the anisotropy through Cr-I-Cr superexchange. In CrCl$_3$ this effect is expected to be much weaker, since the spin-orbit coupling strength varies as the fourth power of the atomic number. Thus, spin-orbit coupling is expected to be negligible for both elements in CrCl$_3$, and this is likely the reason for the much smaller anisotropy field observed in CrCl$_3$ than CrI$_3$.

The low magnetic anisotropy and ferromagnetic in-plane order make CrCl$_3$ potentially very attractive for van der Waals heterostructure studies. Figure \ref{fig:PD}b shows the magnetic structure of two adjacent layers of CrCl$_3$ in zero field, a relatively small out-of-plane field, and a relatively small in-plane field. These distinct magnetic orderings would produce distinct magnetic proximity effects on neighboring materials. Importantly, tuning among these and any intermediate magnetic states should be possible, which would enable exquisite control over interactions and associated functionalities when coupled to appropriate electronic or optical materials within a heterostructured device.

Finally, we show here that monolayer CrCl$_3$ can indeed be realized by exfoliation of bulk crystals. Figure \ref{fig:layers}a shows an AFM image of a flake of CrCl$_3$ on a SiO$_2$ substrate. Several step edges are observed across the flake. Figure \ref{fig:layers}b shows the height profile across an edge measured along the path indicated by the red line in Figure \ref{fig:layers}a. The single layer spacing in CrCl$_3$ determined from the bulk crystal structure is $c~\rm{sin}\beta = 0.58$\,nm, indicated on the figure. The measured step height of $\approx 0.6$\,nm indicates that this step is one monolayer high. By correlating changes in optical contrast with the step height measured using AFM, optical contrast can be used to qualitatively assign layer number.  An optical micrograph of an ultrathin specimen cleaved from a bulk crystal is shown in Figure \ref{fig:layers}c on a 90\,nm SiO$_2$ substrate. Measured optical contrast is used to identify the thicknesses of different regions, which is shown on the image in units of CrCl$_3$ layers. The inset shows a separate specimen determined to be one monolayer thick. It is expected that the single-layer CrCl$_3$ may display ferromagnetism at low temperature, as observed in CrI$_3$ \cite{Huang-2017} and as predicted by first principles calculations for all of the chromium trihalides \cite{Liu-2016, Zhang-2015}.

\begin{figure}
\begin{center}
\includegraphics[width=3.4in]{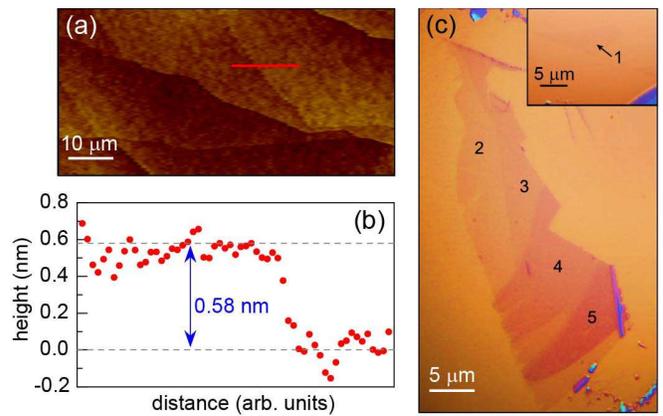}
\caption{\label{fig:layers}
(a) AFM micrograph showing several step edges on a cleaved surface of a CrCl$_3$ crystal. (b) Height profile across the step edge indicated by the red line in (a), with the CrCl$_3$ layer spacing of 0.58\,nm determined from the bulk crystal structure \cite{Morosin-1964} indicated on the plot. (c) Optical micrograph of an ultrathin specimen cleaved from a bulk CrCl$_3$ crystal. Regions of varying optical contrast are labeled by the corresponding thickness in number of CrCl$_3$ layers. A separate monolayer specimen is shown in the inset.
}
\end{center}
\end{figure}

\section{Summary and Conclusions}

We have presented a thorough investigation of the bulk properties of the layered antiferromagnet CrCl$_3$. The magnetic behavior of our crystals is in general agreement with literature reports, indicating ferromagnetic correlations emerging upon cooling before entering the long range antiferromagnetically ordered ground state. Low magnetic anisotropy is observed in the ordered state. A spin flop is seen near 150\,Oe when the field is applied in plane, and fields of several kOe are sufficient to polarize the moments in any direction. In addition, we find that the magnetic susceptibility shows a small anomaly at the crystallographic transition temperature, indicating some coupling of the magnetism to the lattice in the paramagnetic state. Spin-lattice coupling is also apparent in our first principles calculations, in which the experimental structure is matched only when magnetism is included in the calculations. Our heat capacity measurements confirm the two-step nature of the magnetic ordering transition. A broad feature appears near 17\,K indicating the development of ferromagnetism and a sharp, lambda-like anomaly is seen at 14\,K as the antiferromagnetic long range order forms. The observed evolution of the heat capacity with applied magnetic field supports this scenario, and allows the construction of the magnetic phase diagram for CrCl$_3$ shown in Figure \ref{fig:PD}. Neutron scattering experiments would be highly desirable to gain further insight into the development of the magnetic order, specifically to elucidate the potential roles of magnetic fluctuations, short range order, or 2 dimensional order. We have demonstrated that cleaving of CrCl$_3$ crystals into stable monolayer specimens is possible, and note several properties that make CrCl$_3$ a promising material for incorporating magnetism into van der Waals heterostructures at low temperatures: bulk crystals are easily grown, stable in air, electrically insulating, and have magnetic order that can be manipulated with relatively low external fields.


\section*{Acknowledgements}

Research sponsored by the US Department of Energy, Office of Science, Basic Energy Sciences, Materials Sciences and Engineering Division, and by the Laboratory Directed Research and Development Program of Oak Ridge National Laboratory, managed by UT-Battelle, LLC, for the U.S. Department of Energy (WMC). Computational resources were provided by the National Energy Research Scientific Computing Center (NERSC), which is supported by the Office of Science of the U.S. DOE under Contract No. DE-AC02-05CH11231.


%

\end{document}